\documentclass[acus]{JAC2000}

\usepackage{graphicx}

\begin{document}
\title{\flushright{THAT002}\\[15pt] \centering ON THE USE OF CORBA IN \\ HIGH LEVEL SOFTWARE 
APPLICATIONS AT THE SLS
}

\author{M. B\"{o}ge, J. Chrin \\
Paul Scherrer Institut, 5232 Villigen PSI, Switzerland \\ 
\phantom{Paul Scherrer Institut, 5232 Villigen PSI, Switzerland}\\ 
\phantom{Paul Scherrer Institut, 5232 Villigen PSI, Switzerland}  
}

\maketitle

\hyphenation{suite Event access Broker applications servers server consumers internals 
persistent App-lications ABS Subjects Template Library object CORBA client DSI SII call
BD stubs Interface activations levels hosts}

\begin{abstract}
Beam dynamics applications at the Swiss Light Source (SLS)
have benefitted from a distributed computing environment
in which the Common Object Request Broker Architecture 
(CORBA) forms the middleware layer and access point to
several different software components.
A suite of remote CORBA server objects provides the 
client with a convenient and uniform interface to
the CDEV (Common DEVice) controls library,  
the TRACY accelerator physics package, 
the Oracle database, and
an event-logging facility. 
Use is made of methods provided by the CORBA Portable 
Object Adaptor for accessing ORB functions, such as object 
activation and object persistence, the Implementation Repository 
for the automatic reactivation of servers, and the CORBA 
Event Service for the propagation of controls and physics data.
An account of the CORBA framework, as used by applications 
in the commissioning and first operation of the SLS, 
is presented. 
\end{abstract}

\section{MOTIVATION}
\begin{figure}[htb]
\vskip 0.05cm
\centering
\includegraphics*[width=82.0mm]{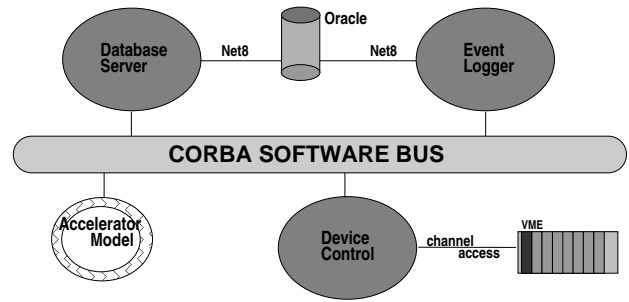}
\caption{DOC components serving BD applications}
\label{fig_1}
\end{figure}

The Swiss Light Source (SLS) is a synchrotron
light source located at the Paul Scherrer
Institute (PSI) in Switzerland. Its most
major component, a 2.4 GeV electron storage
ring, was recently commissioned and is now
delivering light of high brilliance to
experimenters from multiple disciplines.
Several high-level beam dynamics (BD) applications
have been developed for the operation
and monitoring of the SLS accelerator 
facilities. 
Fig.~\ref{fig_1} captures typical
components required by BD applications.
Their number and demand on 
computer resources motivated, in part, 
a desire for a distribued computing environment. 
To this end, the Common Object Request Broker (CORBA),
an emerging standard for distributed
object computing (DOC), has been employed. 
Its use at the SLS has allowed us to realize
the potential benefits of distributed computing 
and to simultaneously exploit features inherent
to CORBA, such as the interoperability between objects 
of different race (language) and creed (platform).
Complex tasks, such as the modeling
of the SLS accelerators, can thus be handled
by dedicated computers, and developed
into reusable components that 
can be accessed 
through remote method invocations.
Perservering with the notion of DOC 
and developing the entire suite
of BD components as  
CORBA objects, further elevates the level
at which applications are designed and implemented.
Platforms hosting high-level software applications
are no longer limited to the libraries and 
extensions available to the host operating system
as the introduction of a CORBA middleware layer
serves to {\it extend} the developers chosen 
programming language. 
BD application 
developers are, henceforth, able to focus on the 
specifics of the application at hand, such as
determining user-friendly graphical
interfaces, rather than struggle with
the intricate internals of numerous application
program interfaces (APIs) and low-level 
communication protocols.
\section{The CORBA Architecture}
The most fundamental component of CORBA is the 
Object Request Broker (ORB) whose task is
to facilitate communication between objects.
Given an Interoperable Object Reference (IOR), 
the ORB is able to locate target objects and transmit 
data to and fro
remote method invocations.
The interface to a CORBA object is 
specified using the CORBA Interface
Definition Language (IDL). 
An IDL compiler translates the IDL definition
into an application programming language,
such as C++ or Java, 
generating IDL stubs and skeletons
that respectively provide the framework
for client-side and server-side proxy code.
Compilation of applications incorporating IDL 
stubs provides a strongly-typed 
Static Invocation Interface (SII).
Conversely,
a more flexible communication mechanism
can be established through the use of
the Dynamic Invocation Interface (DII)
and the Dynamic Skeleton Interface (DSI)
allowing objects to be created without
{\it prior} knowledge of the IDL interface.
In such cases, a description of the interface
is retrieved at runtime from an Interface Repository (IFR),
a database containing the pertinent meta-data. 
Requests and responses between objects
are delivered in a standard format defined 
by the Internet Inter-ORB Protocol (IIOP),
a communications protocol which 
adheres to the CORBA General Inter-ORB Protocol
(GIOP) specification and, as such, 
acts as the base for CORBA interoperability on 
the internet.
Requests are marshaled in a platform
independent format,
by the client stub (or in the DII),
and unmarshaled on the server-side 
into a platform specific format by the
IDL skeleton (or in the DSI) and the
object adaptor, which serves as a mediator 
between an object's implementation, 
the servant, and its ORB, 
thereby decoupling user code 
from ORB processing.
In its mandatory version, the 
Portable Object Adaptor (POA)
provides CORBA objects with a common set
of methods for accessing ORB functions,  
ranging from user authentication to object 
activation and object persistence.
It's most basic task, however,
is to create object references and to
dispatch ORB requests aimed at target objects 
to their respective servants. 
The characteristics of the POA
are defined at creation 
time by a set of POA policies.
A server can host any number of
POAs, each with its own set of policies
to govern the processing of requests.
Among the more advanced features of the
POA is the servant manager 
which assumes the role of reactivating
server objects (servants) as they are required.
It also provides a mechanism to save and restore 
an object's state.
This, coupled with the use of the
Implementation Repository (IMR), 
that handles
the automated start and restart of servers,
realizes object persistency.
Requests for server reactivation can,
alternatively, be 
delegated to a single default servant 
which provides implementations for many objects, 
thereby increasing the
scalability for CORBA servers.
\begin{figure}[htb]
\vskip 0.05cm
\centering
\includegraphics*[width=82.0mm]{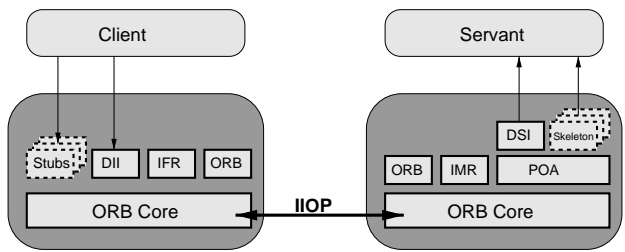}
\caption{The CORBA client-server architecture}
\label{fig_orb}
\vskip 0.04cm
\end{figure}

Fig.~\ref{fig_orb} shows the components
of the CORBA architectural model.
The ORB core is implemented as a runtime
library linked into client-server applications. 

\subsection{Client and Server Perspectives}
Despite the plethora of new terms and concepts
introduced,
CORBA, nevertheless,
remains true to 
the DOC objective of providing
developers with familiar object-oriented
techniques with which to program 
distributed applications. 
Indeed, from the client perspective, 
once an IOR 
is obtained (typically from a Naming Service
which maps names to object references)
a remote method invocation 
essentially takes on the welcoming
appearance of a local function call:
\texttt{\small object-$>$operation(arguments);}
%
%
%
%
whilst the communication details of 
client-server programming 
are essentially hidden from the client,
a more intimate involvement with the ORB
is required when developing servers.
In particular appropriate POA policies need 
to be chosen to configure object adaptors
that best fufill the requirements of the
server.
\subsection{Power to the POA}
Transient and persistent objects
are two categories of objects that relate to the
lifespan policies of the POA.
A transient object is short-lived with a lifetime
that is bounded by the POA in which it was created.
A persistent object, on the other hand, is long-lived 
with a lifetime that is unbounded. It can consequently
outlive the very server process wherein it was created.
This has several advantages.
A server may be shutdown whenever it is not
needed to save resources. Server updates can be
implemented transparantly by restarting the server.
In developing a DOC environment, the command to start 
a server may be replaced with a remote shell invocation
and the next server instance run remotely, without
the client being aware. Persistent objects also
maintain their identify after a server crash.
Whilst the POA supports and implements persistent objects,
it does not handle the administrative aspects of server 
activations.
This is managed by the IMR which
stores an activation 
record for each server process; it is consulted 
automatically whenever a (re-)launch of a server is mandated.
\begin{figure}[htb]
\vskip 0.05cm
\centering
\includegraphics*[width=82.0mm]{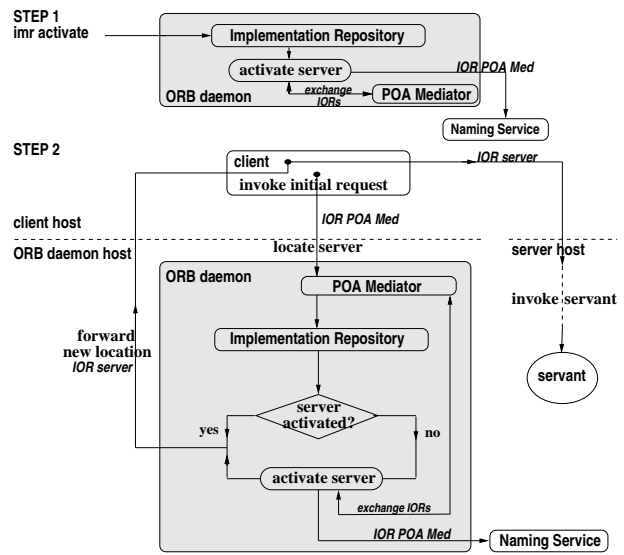}
\caption{Server activation through the IMR}
\label{fig_imr}
\vskip 0.04cm
\end{figure}

Fig.~\ref{fig_imr} illustrates the role of the IMR
in the (re-)activation of servers. 
The first instance of the server is started by an
administrative procedure (imr activate) and object 
references, 
pointing to the POA Mediator within the ORB daemon process, 
are exported to the Naming 
Service \mbox{(step~1)}.
The ORB daemon listens
for CORBA client connection attempts and
assists the client in connecting to its destined server.
This is done through the POA Mediator whose task is
to intercept initial client requests, 
to (re-)activate the server if so required, and to
forward the actual server location 
to the client for all subsequent operations  
(step 2).
Thus, by virtue of the capabilities of the POA,
and the activation techniques of the IMR,
BD applications are never starved of the
servers they require.
 
\subsection{The Event Service}
A reactive, event-based, form of programming is 
also supported by the CORBA Event Service 
which provides services for the creation 
and management of CORBA event channels.  
These may be used by CORBA
supplier/consumer clients to propagate events asynchronously on a
push or pull basis. Event channels are created and registered with
the CORBA Naming Service allowing clients to obtain object
references in the usual manner. Communication is anonymous in that
the supplier does not require knowledge of the receiving consumers.
Publicized inadequacies of the Event Service are a lack of
explicit quality of service (QoS) control, the necessity of
propagating event data with type {\it CORBA::any}, and the absence of
event filtering. Nevertheless, by applying 
a few simple design techniques, 
these limitations can be largely circumvented and
the CORBA Event Service has been usefully employed in the monitoring
of hardware devices and in the distribution of recalibrated data 
to client consumers.
The CORBA Event Service is ultimately to be replaced by
the CORBA Notification Service which systematically 
addresses the shortcomings of the Event Service.
\section{Server Synopsis}

Server objects, typically of persistent type, 
have been developed using the CORBA product
MICO~\cite{ref_mico01}, a fully compliant 
implementation of the CORBA 2.3 standard.
The services these objects provide are briefly 
reported here. An expanded description,  
together with the specifications of the hardware
and system components of the server hosts,
appears elsewhere~\cite{ref_pcapac00}.


\subsection{The Accelerator Model}
A dedicated host runs the servers
that perform the computer intensive 
modeling of the SLS accelerators.
Procedures utilise the complete 
TRACY accelerator physics library, 
enabling clients to employ 
accelerator optimization
procedures {\it online}.

\subsection{Device Controls}
The CDEV C++ class controls library 
provides the API to the EPICS-based 
accelerator device control system. 
The CDEV server supplies 
functionality for both synchronous and 
asynchronous interactions with the control system.
Monitored devices and recalibrated data 
are propagated to clients through CORBA event channels.

\subsection{Making a Statement with CORBA}
A database server provides access to
Oracle instances through the Oracle Template
Library (OTL) and the Oracle Call Interface (OCI).
Methods executing  
SQL statements that perform database retrieval and 
modification operations have been provided.
Interestingly, database access through the 
CORBA interface ({\it with} data marshaling)
takes half the time than that 
through the JDBC API.

\subsection{Monitoring Servers and Applications}
A CORBA message server has been developed
using the the UNIX syslog
logging facility, profiting directly
from the reliability of standard
UNIX services. 
Run-time messages are sent to the logger with
various priority levels, the threshold
for which can be adjusted dynamically
for any given servant.
This is particularly useful
during the development stage, where 
for instance, debugging
can be activated  
without the need to recompile.


\section{SLS STORAGE RING OPERATION}
Several applications, written mainly in
Tcl/Tk or Java, 
have been successfully introduced 
for the commissioning
and operation of the SLS booster and
storage rings \cite{ref_pac01},
making ample use of the CORBA
framework provided. Server objects were
extensively tested through 
invocations initiated by a variety of
client processes. Operator intervention was 
minimal with clients able to interact 
spontaneously with the many servers and to 
  display their event data.
This is exemplified 
by the slow global orbit feedback system
(3 Hz sampling rate),
which is both a consumer to event generated data 
and a party to remote methods invocations on a 
variety of servers. 
A fast version, for which dedicated
low-level hardware is to be installed
(4 kHz sampling rate),
will challenge the reaches of our model \cite{ref_abs01}.


\section{CONCLUSION}
The CORBA middleware has
served to extend the capabilities
of the programming languages used 
by BD application developers,
thereby elevating the level
at which high-level software 
applications are designed and 
implemented.
The power and flexibility of the POA,
coupled with the server activation records
stored within the IMR, has been exploited to
provide a robust and modular 
CORBA based client-server framework.
The model has been proved to be both
reliable and stable by the 
many applications deployed
in the commissioning
and first operation of the SLS.

\end{document}